\newif\ifmulticol	\multicoltrue
\newif\ifshowgit	\showgittrue		
\newif\ifgitlocal	\gitlocalfalse		
\newif\ifbiblatex	\biblatexfalse		
\newif\ifbibnum		\bibnumtrue 		
\newif\iflineno		\linenofalse
\newif\iftoc		\tocfalse

\newif\iflucida		\lucidafalse
\newif\ifcm			\cmtrue
\newif\iflibertine	\libertinefalse
\newif\ifcharter	\charterfalse

\newcommand*{\secnumdepth}{0}			
\newcommand*{\tocdepth}{2}				
\newcommand*{\reftitle}{References}		
\newcommand*{\mytitleskip}{-0.2in}		

\multicoltrue\showgittrue\gitlocaltrue\biblatexfalse\bibnumtrue

\newcommand*{\mydocfontsize}{\ifcharter11pt\else\iflibertine11pt\else10pt\fi\fi}
\renewcommand*{\mydocfontsize}{11pt}
\newcommand*{\setcol}{\ifmulticol twocolumn\else onecolumn\fi}

\documentclass[\mydocfontsize,\setcol]{article}

\newcommand*{\pdftitle}{Simple unity among the fundamental equations of science}

\input price.sty

\newcommand{\bdq}{\GD\mathbf{q}}
\newcommand{\bq}{\mathbf{q}}
\newcommand{\dbq}{\dd\bq}

\newcommand{\Dq}{\GD q}
\newcommand{\bda}{\GD\mathbf{a}}
\newcommand{\ba}{\mathbf{a}}

\newcommand{\bw}{\mathbf{w}}
\newcommand{\obq}{\dbq}
\newcommand{\oq}{{\dot{q}}}

\newcommand{\bdz}{\GD\mathbf{z}}
\newcommand{\bz}{\mathbf{z}}

\newcommand{\bmm}{\mathbf{m}}

\newcommand{\zbar}{\bar{z}}

\newcommand{\abar}{\bar{a}}

\newcommand{\br}{\mathbf{r}}
\newcommand{\obr}{\dd\br}

\newcommand{\bF}{\mathbf{F}}
\newcommand{\bI}{\mathbf{I}}

\newcommand{\Gfb}{\boldsymbol{\Gf}}

\newcommand{\KL}[2]{\mathcal D\left(#1||#2\right)}
\newcommand{\D}{\mathcal D}
\newcommand{\J}{\mathcal J}
\newcommand{\F}{\mathcal F}

\newcommand{\cov}{\mathrm{Cov}}

\newcommand{\GDF}{\GD_{\raisebox{-2pt}{$\scriptstyle \mathrm{F}$}}}

\newcommand{\dlog}{\dd\log}
\newcommand{\bL}{\mathbf{L}}

\newcommand{\qt}{q_\Gth}

\newcommand*{\GD}{\Delta}

\newcommand*{\Gom}{\omega}

\newcommand*{\Gth}{\theta}
\newcommand*{\Gf}{\phi}

\DeclarePairedDelimiter\abs{\lvert}{\rvert}
\DeclarePairedDelimiter\norm{\lVert}{\rVert}
\DeclarePairedDelimiter\angb{\langle}{\rangle}
\DeclarePairedDelimiter\lrb{\lbrack}{\rbrack}
\DeclarePairedDelimiter\lr{\lparen}{\rparen}

\makeatletter
\let\oldabs\abs \def\abs{\@ifstar{\oldabs}{\oldabs*}}
\let\oldnorm\norm \def\norm{\@ifstar{\oldnorm}{\oldnorm*}}
\let\oldangb\angb \def\angb{\@ifstar{\oldangb}{\oldangb*}}
\let\oldlrb\lrb \def\lrb{\@ifstar{\oldlrb}{\oldlrb*}}
\let\oldlr\lr \def\lr{\@ifstar{\oldlr}{\oldlr*}}
\makeatother

\DeclareMathOperator{\E}{E}
\newcommand*{\dd}{\textrm{d}}

\newcommand*{\Eq}[1]{eqn~\ref{eq:#1}}

\newcommand*{\Fig}[1]{Fig.~\ref{fig:#1}}

\newcount\BoxNum \BoxNum 1\relax
\makeatletter
\newcommand*{\boxlabel}[1]{%
  \protected@write \@auxout {}{\string \newlabel {box:#1}{{\the\BoxNum}}{}}%
  \advance\BoxNum 1\relax}
\makeatother

\usepackage{booktabs}

\newcommand*{\mytitle}{\pdftitle}

\newcommand*{\myauthor}{Steven A.\ Frank}
\newcommand*{\myaddress}{Department of Ecology and Evolutionary Biology, University of California, Irvine, CA 92697--2525, USA}
\newcommand*{\mycontact}{\href{https://stevefrank.org}{https://stevefrank.org}}

\newcommand*{\mykeywords}{Price equation, invariance, conservation laws, information theory, physical mechanics}

\setabstract{-0.07}{\iftoc-2.0\else-0.02\fi}{%
The Price equation describes the change in populations. Change concerns some value, such as biological fitness, information or physical work. The Price equation reveals universal aspects for the nature of change, independently of the meaning ascribed to values. By understanding those universal aspects, we can see more clearly why fundamental mathematical results in different disciplines often share a common form. We can also interpret more clearly the meaning of key results within each discipline. For example, the mathematics of natural selection in biology has a form closely related to information theory and physical entropy. Does that mean that natural selection is about information or entropy? Or do natural selection, information and entropy arise as interpretations of a common underlying abstraction? The Price equation suggests the latter. The Price equation achieves its abstract generality by partitioning change into two terms. The first term naturally associates with the direct forces that cause change. The second term naturally associates with the changing frame of reference. In the Price equation's canonical form, total change remains zero because the conservation of total probability requires that all probabilities invariantly sum to one. Much of the shared common form for the mathematics of different disciplines may arise from that seemingly trivial invariance of total probability, which leads to the partitioning of total change into equal and opposite components of the direct forces and the changing frame of reference.
}

\begin{document}

\mymaketitle

\iftoc\mytoc{-24pt}{\newpage}\fi

\section{Introduction}

Problems often concern change. How does natural selection alter a population? How does force change position and velocity? How does climate affect biodiversity?

Many forces act on a system. We cannot know all of them. The Price equation helps by partitioning change into components. One component isolates particular forces. The second component includes everything else. 

For a system that changes in a specific way, the equation does not tell us which forces to isolate. We could focus on climate. Or we could focus on a meteor explosion. No matter what forces we isolate, the overall system changes in the same way. 

The Price equation's isolation of forces helps because it ``focuses attention on the forces, not on the moving body'' \autocite{lanczos86the-variational}. Force associates with cause. We may be more interested in understanding cause rather than describing motion.

Different fields focus on different forces. Yet, no matter the focus, disciplines often share a common underlying expression of force and a common partitioning of total change into components. 

That unity arises from a simple conservation law for total change. The Price equation partitions the conserved total into a component for the direct forces and a component for the changing frame of reference. 

From the Price equation's abstract partitioning, one obtains many of the fundamental mathematical equations of change that recur across disciplines. For example, common expressions of information arise as a simple invariant measurement scale of geometric divergence caused by the forces of change.

In prior articles, I used the Price equation to show the unity among the fundamental equations of many different fields of science \autocite{frank18the-price,frank17universal}. This article emphasizes the basic concepts and geometric intuition\footnote{This article is an adaptation of a prior publication \autocite{frank18the-price}.}.

\newcommand*{\trow}[3]{$#1$& \parbox[t]{10.5cm}{\raggedright\hangindent=0.1pt\hangafter=1 #2}&~(\ref{eq:#3})\\[3pt]}
\newcommand*{\ts}{\vskip4pt}
\newcommand*{\tbi}[1]{\textbf{\textit{#1}}}

\begin{table*}[t]
\caption{Definitions of key symbols and concepts.}
\label{tbl:symbols}
\centering
\begin{tabular}{clc}
\toprule
\textbf{Symbol}	& \textbf{Definition}	& \textbf{Equation}\\
\midrule

\trow{\bq}{Vector of frequencies with $\sum q_i=1$}{price}
\trow{\bz}{Values with average $\zbar=\bq\cdot\bz$; use $\bz\equiv\ba,\bF$, etc.\ for specific interpretations}{price}
\trow{\bdq}{Discrete changes, $\GD q_i=q_i'-q_i$, may be large}{price}
\trow{\obq}{Small, differential changes, $\bdq\rightarrow \obq$}{dlogq}
\trow{\ba}{Relative change of the $i$th type, $a_i=\GD q_i/q_i\rightarrow\oq_i/q_i=\log q_i'/q_i$}{adef}
\trow{\bmm}{Malthusian parameter, $\bmm=\log \bq'/\bq$, log of relative fitness, $\bw$}{malthus}
\trow{\bw}{Relative fitness, $w_i=q_i'/q_i$, with $\bmm=\log \bw$}{adef}
\trow{\bF}{Direct nondimensional forces, may be used for values $\bz\equiv\bF$}{work}
\trow{\bI}{Inertial nondimensional forces, may be interpreted as acceleration (\ref{eq:Ilog})}{Idef}
\trow{\bdq\cdot\bF}{Abstract notion of physical work as displacement multiplied by force}{work}
\trow{\F}{Fisher information, nondimensional expression}{fisherDist}
\trow{\bL}{Likelihoods, $L_\Gth$, for parameter values, $\Gth$; interpreted as force, $\bF\equiv\bL$}{logL}
\trow{\GDF}{Partial change caused by direct forces, e.g., $\bdq\cdot\bF$ or $\bdq\cdot\Gfb$ or $\bdq\cdot\bL$}{work}
\trow{\norm{\cdot}}{Euclidean vector length, e.g., $\norm{\bz}$ or $\norm{\bF}$ or $\norm{\bdq}$}{normdef}
\trow{\br}{Unitary coordinates, $\br=\sqrt{\bq}$, with $\norm{\br}=1$ as invariant total probability}{unitary}

\bottomrule
\end{tabular}
\end{table*}

\section{The Price equation}

I begin with an abstract mathematical derivation of the Price equation. I then connect the abstract mathematical equations of change to particular examples. 

The examples in later sections include the work done by physical forces, the information gained by the force of natural selection, and the updated Bayesian inference achieved by the ``force'' of new data. See Frank \autocite{frank18the-price} for further examples.

The Price equation describes the change in the average value of some property between two populations \autocite{price72extension,frank12naturalb}. A population is a set of things. Each thing has a property indexed by $i$. Those things with a common property index comprise a fraction, $q_i$, of the population and have average value, $z_i$, for whatever we choose to measure by $z$. See Table \ref{tbl:symbols} for a summary of notation.

Write $\bq$ and $\bz$ as the vectors over all $i$. The population average value is $\zbar=\bq\cdot\bz=\sum q_iz_i$, summed over $i$.

A second population has matching vectors $\bq'$ and $\bz'$. Here, $q_i'$ is the fraction of the second population derived from entities with index $i$ in the first population. Similarly, $z_i'$ is the average value in the second population of members derived from entities with index $i$ in the first population. Let $\GD$ be the difference between the derived population and the original population, $\GD\bq=\bq'-\bq$ and $\GD\bz=\bz'-\bz$.

The difference in the averages is $\GD\zbar=\bq'\cdot\bz'-\bq\cdot\bz$. By using the definitions for $\GD\bq$ and $\GD\bz$, we can write the change in the average as the abstract form of the Price equation
\begin{equation}\label{eq:price}
  \GD\zbar=\bdq\cdot\bz+\bq'\cdot\bdz.
\end{equation}
The first term, $\bdq\cdot\bz$, is the partial difference of $\bq$ holding $\bz$ constant. The second term, $\bq'\cdot\bdz$, is the partial difference of $\bz$ holding $\bq$ constant. In the second term, we use $\bq'$ as the constant value because, with discrete differences, one of the partial change terms must be evaluated in the context of the second set.

Note that $\bq$ has a clearly defined meaning as frequency, whereas $\bz$ may be chosen arbitrarily as any values assigned to members. The values, $\bz$, define the frame of reference. Because frequency is clearly defined, whereas values are arbitrary, the frequency changes, $\bdq$, take on the primary role in analyzing the structural aspects of change that unify different subjects. 

The primacy of frequency change naturally labels the first term, with $\bdq$, as the changes caused by the direct forces acting on populations. Because $\bq$ and $\bq'$ define a sequence of probability distributions, the primary aspect of change concerns the dynamics of probability distributions.

The arbitrary aspect of the values, $\bz$, naturally labels the second term, with $\bdz$, as the changes caused by the forces that alter the frame of reference. Those forces that change the frame of reference are sometimes called the inertial forces \autocite{lanczos86the-variational}. 

It is, of course, possible to interpret the terms in other ways. The equation itself is a simple mathematical identity. That identity has no intrinsic meaning beyond the fundamental partitioning of the change in an average value, $\GD\zbar$, into two components of change.

The Price equation is often written as
\begin{equation*}
  \GD z=\cov(w,z)+\E(w\GD z),
\end{equation*}
in which $w$ is relative fitness, as defined below \Eq{adef}. This expression is equivalent to \Eq{price}, because $\cov(w,z)=\bdq\cdot\bz$ and $\E(w\GD z)=\bq'\cdot\bdz$. The ``$\E$'' means the expectation or average. I focus on the form in \Eq{price}, with vectors and dot products. That form emphasizes the geometry, with natural interpretations in terms of force, distance and constraints imposed by conserved quantities. The commonly used statistical expressions are alternative notations for the same fundamental aspects of distances and geometry \autocite{frank12naturalb}.

\section{Conservation of total probability}

Probabilities are typically normalized to sum to one, $\sum q_i=\sum q_i'=1$. That normalization is usually thought of as a trivial aspect of the definition of probability. However, the conservation of total probability profoundly shapes the form of fundamental equations. Seemingly different subjects often share common expressions because they share invariant total probability and its geometric consequences. The Price equation reveals that unity.

To describe the conservation of total probability in the Price equation, define
\begin{equation}\label{eq:adef}
  a_i=\frac{\Dq_i}{q_i}=\frac{q_i'}{q_i}-1=w_i-1.
\end{equation}
In biology, $w_i=q_i'/q_i$ may be interpreted as relative fitness. 

In the Price equation, we can use any value for $\bz$. Let $\bz\equiv\ba$. The Price equation becomes
\begin{equation}\label{eq:canonical}
  \GD\abar=\bdq\cdot\ba+\bq'\cdot\bda=0.
\end{equation}
The equality to zero follows because the average of $\ba$ is always zero: $\abar=\sum q_ia_i=\sum\Dq_i=0$. All of the changes in probability, $\Dq_i$, must sum to zero in order to keep the total probability constant at one.

We can first study universal aspects of the canonical invariant form based on $\ba$ in \Eq{canonical}. We can then derive broader results by simply making the coordinate transformation $\ba\mapsto\bz$, yielding the most general expression of the abstract Price equation in \Eq{price}.

\begin{figure*}[t]
\centering
\includegraphics[width=15.8cm]{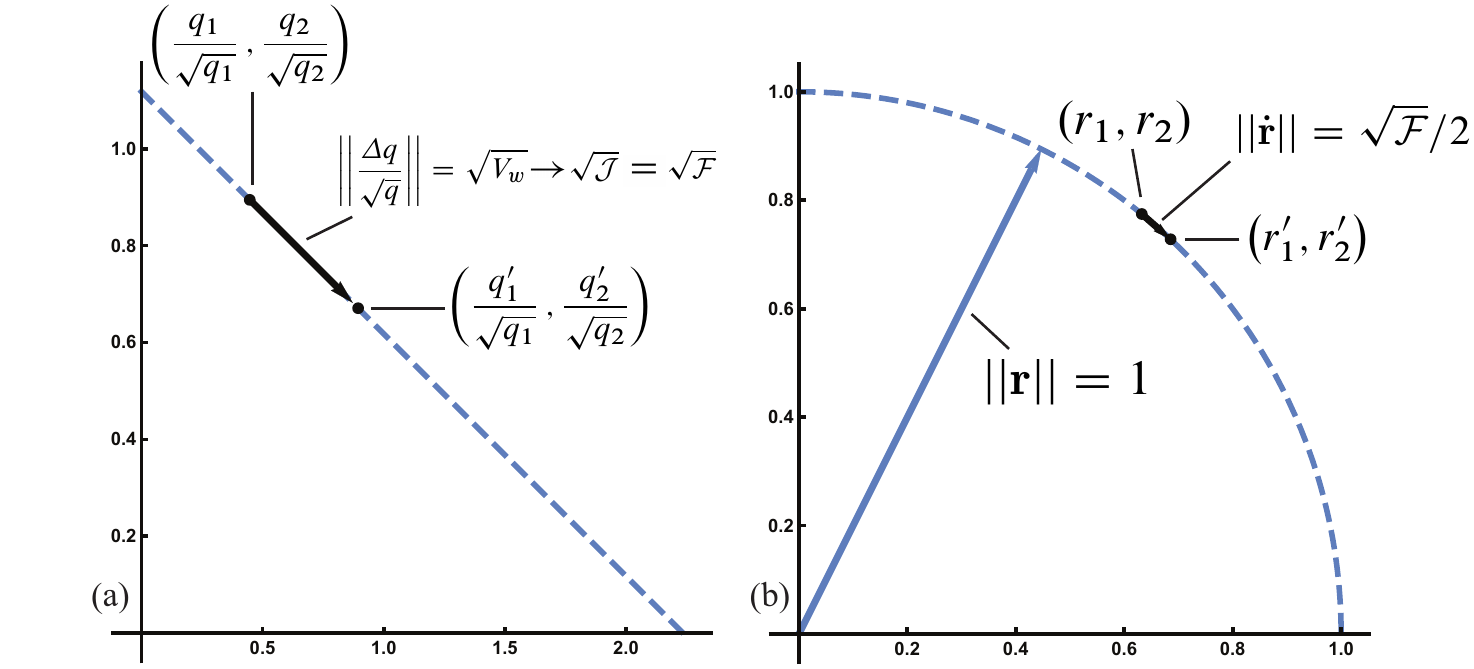}
\caption{Geometry of change by direct forces. (\textbf{a}) The abstract physical work of the direct forces as the distance moved between the initial population with probabilities, $\bq$, and the altered population with probabilities, $\bq'$. For discrete changes, the probabilities are normalized by the square root of the probabilities in the initial set. The distance can equivalently be described by the various expressions shown, in which $V_w$ is the variance in fitness from population biology, $\J$ is the Jeffreys divergence from information theory, and $\F$ is the Fisher information metric which arises in many disciplines. (\textbf{b}) When changes are small, the same geometry and distances can be described more elegantly in unitary square root coordinates, $\br=\sqrt{\bq}$, with $\mathbf{\dot{r}}\equiv\dd\br$. The symbol ``$\rightarrow$'' denotes the limit for small changes. From Frank \autocite{frank18the-price}.}
\label{fig:geometry}
\end{figure*}   

\section{Distance, force and work}

The first term of the Price equation expresses an abstract notion of physical work. The nondimensional work is the product of a force acting on an entity multiplied by the distance that the entity moves. 

For example, consider natural selection as a force. The strength of natural selection multiplied by the distance moved by the population is the work accomplished by natural selection. 

When we think about distance, force and work abstractly, we obtain general insight. Natural selection and other problems arise as special cases. To see that generality, begin by writing the standard Euclidean geometry vector length as the square root of the sum of squares
\begin{equation}\label{eq:normdef}
  \norm{\bz}=\sqrt{\sum z_i^2}.
\end{equation}

For any vector $\bz$, the first term of the Price equation is
\vspace{12pt}
\begin{equation*}
  \bdq\cdot\bz=\norm{\bdq}\norm{\bz}\cos\Gom,
\end{equation*}
in which $\Gom$ is the angle between the vectors $\bdq$ and $\bz$. If we interpret $\bz\equiv\bF$ as an abstract, nondimensional force, then
\begin{equation}\label{eq:work}
   \GDF\mskip1mu\zbar=\bdq\cdot\bF=\norm{\bdq}\norm{\bF}\cos\Gom
\end{equation}
expresses an abstract notion of work as the distance moved, $\norm{\bdq}$, multiplied by the component of force acting along the path of motion, $\norm{\bF}\cos\Gom$. 

This expression for work arises in the first term of the Price equation as the partial change in response to the direct forces, $\GDF\mskip1mu\zbar$. 

\section{Geometry of change in populations}

\subsection{Divergence between populations}

If we let $\bz\equiv\ba$ describe the relative growth of the various probabilities, $a_i=\GD q_i/q_i$, then the divergence between populations caused by the directly acting forces can be expressed as 
\begin{equation}\label{eq:diverge}
  \GDF\mskip1mu\abar=\bdq\cdot\ba
    =\sum\lr{\frac{\GD q_i}{\sqrt{q_i}}}^2=\norm{\frac{\bdq}{\sqrt{\bq}}}^2
    =V_w.
\end{equation}
If we choose to interpret $\ba$ as an abstract notion of force, or fitness, acting on frequency changes, then $\bdq\cdot\ba$ is the work, with magnitude $\norm{\bdq\;/\mskip-1mu\sqrt{\bq}\,}^2$, that separates the probability distribution $\bq'$ from $\bq$. In this article, the division of vectors, such as $\bdq\;/\mskip-1mu\sqrt{\bq}\,$, means elementwise division.

That value of work is equal to the variance in fitness, $V_w$. From \Eq{adef}, $a_i=w_i-1$. Thus, $\bdq\cdot\ba=\cov(w,a)=V_w$. The variance in fitness simply describes the geometric divergence between populations caused by the force of natural selection. See \Fig{geometry}a.

\subsection{Small changes, paths and logarithms}

The Price equation provides an exact description for large, discrete changes. Small, continuous changes are included as a special case. 

In prior articles, I developed the theory fully for discrete changes. In the remainder of this article, I focus on small, continuous changes. That focus on continuity makes the mathematics simpler and highlights conceptual aspects more clearly.

If we think of the separation between populations as a sequence of small changes along a path, with each small change as $\bdq\rightarrow\obq$. This notation means that as the changes, $\bdq$, approach zero, we write those changes in differential notation, $\obq$. With that notation
\begin{equation}\label{eq:dlogq}
  \ba\rightarrow\frac{\obq}{\bq}=\dd\log\bq.
\end{equation}
With the differential notation, the partial change by the direct forces separates the probability distributions of the two populations by the path length
\begin{equation}\label{eq:fisherDist}
  \GDF\mskip1mu\abar=\bdq\cdot\ba\rightarrow\obq\cdot\ba
  	=\norm{\frac{\obq}{\sqrt{\bq}}}^2=\F,
\end{equation}
in which $\F$ is an abstract, nondimensional expression of the Fisher information distance metric \autocite{cover91elements}.

\subsection{Unitary geometric coordinates}

Let $\br=\sqrt{\bq}$. Then $\norm{\br}=1$, expressing the conservation of total probability as a vector of unit length, in which all possible probability combinations of $\br$ define the surface of a unit sphere.

The unitary coordinates, $\br$, also provide a direct description of Fisher information path length as a distance between two probability distributions
\begin{equation}\label{eq:unitary}
  4\norm{\obr}^2 = 4\norm{\dd\sqrt{\bq}}^2
  	=\norm{\frac{\obq}{\sqrt{\bq}}}^2=\F.
\end{equation}
The constraint on total probability makes square root coordinates the natural system in which to analyze Euclidean distances, which are the sums of squares. See Figure~\ref{fig:geometry}b. 

\section{Direct forces and frame of reference}

\subsection{Direct and inertial forces}

For small changes, we can write the canonical Price equation for the conservation of total probability in \Eq{canonical} as
\begin{equation*}
  \dd\abar=\dbq\cdot\ba+\bq\cdot\dd\ba=0.
\end{equation*}
To emphasize the first term as the direct forces acting on frequency change and the second term as the inertial forces that change the frame of reference, write
\begin{equation}\label{eq:dalembert}
  \lr{\bF+\bI}\cdot\dbq=0.
\end{equation}
The first term describes the direct forces
\begin{equation}\label{eq:Flog}
  \bF\equiv\ba=\frac{\dbq}{\bq}=\dlog\bq,
\end{equation}
as in \Eq{dlogq}.
The second term describes the inertial forces
\begin{equation}\label{eq:Idef}
  \bI=\frac{\bq}{\dbq}\,\dd\ba=\frac{\dd\ba}{\ba}=\dlog\ba,
\end{equation}
in which multiplication and division of vectors is elementwise, and $\dlog$ is an operator acting on nonzero quantities that maps an argument $x$ to $\dd x/x$. (Note that $\dlog$ maps its argument to its outcome in a single step, rather than as a logarithm and then a differential. Thus, $a$ can be negative here.) This expression for the inertial forces can be expanded as
\begin{equation}\label{eq:Ilog}
  \bI=\dlog\ba=\dlog\lr{\dlog \bq}=\dlog^2\bq.
\end{equation}
The relative differential, $\dlog$, describes relative change. The second relative differential, $\bI=\dlog^2\bq$, describes the relative acceleration in frequency changes. Thus, the inertial forces acting on the frame of reference can be related to an acceleration. 

\subsection{D'Alembert's principle}

Substituting the expressions for the direct and inertial forces by the relative change and the relative acceleration of frequencies yields
\begin{equation}\label{eq:dalembertLog}
  \lr{\bF+\bI}\cdot\dd\bq=\lr{\dlog\bq+\dlog^2\bq}\cdot\dd\bq=0.
\end{equation}
When written in this form, the canonical Price equation of \Eq{canonical} is an abstract, nondimensional generalization of d'Alembert's principle for probability distributions that conserve total probability \autocite{frank18the-price,frank17universal,frank15dalemberts}. 

D'Alembert's principle is a fundamental expression of physical mechanics \autocite{lanczos86the-variational}. The principle generalizes Newton's second law, force equals mass times acceleration. In one dimension, Newton's law is $F=-mI$, for force, $F$, and mass, $m$, times acceleration, $-I$. In my abstract nondimensional expressions, $m$ drops out, so that $F+I=0$. 

D'Alembert generalizes Newton's law to a statement about motion in multiple dimensions such that, in conservative systems, the total work for a displacement, $\dbq$, and total forces, $\bF+\bI$, is zero. Work is the distance moved multiplied by the force acting in the direction of motion. 

In terms of the canonical Price equation with conserved total probability, the change of a probability distribution between two populations can be partitioned into the balancing work components of the direct forces, $\dbq\cdot\bF$, and the inertial forces, $\dbq\cdot\bI$. We can often specify the direct forces in a simple and clear way. The balancing inertial forces may then be analyzed by d'Alembert's principle \autocite{lanczos86the-variational}.

\subsection{Frame of reference}

Here is a simple intuitive description of d'Alembert's principle \autocite{wikipedia15fictitious}. You are sitting in a car at rest, and the car suddenly accelerates. You feel thrown back into the seat. But, even as the car gains speed, you effectively do not move in relation to the frame of reference of the car: Your velocity relative to the car remains zero. That net zero velocity can be thought of as the balance between the direct force of the seat pushing on you and the inertial force sending you back as the car accelerates forward.

As long as your frame of reference moves with you, then your net motion in your frame of reference is zero. Put another way, there is a changing frame of reference that zeroes net change by balancing the work of the direct forces against the work of the inertial forces. Although the system is a dynamic expression of changing components, it also has an overall static, equilibrium quality that aids analysis. As Lanczos \autocite{lanczos86the-variational} emphasizes, d'Alembert's principle ``focuses attention on the forces, not on the moving body $\ldots$''. (These two paragraphs are from Frank \autocite{frank17universal}.)

\subsection{Conservative and nonconservative systems}

From \Eq{fisherDist}, the work of the direct forces, $\obq\cdot\bF=\F$, is the Fisher information path length that separates the probability distributions, $\bq'$ and $\bq$. The inertial forces cause a balancing loss, $\obq\cdot\bI=-\F$, which describes the loss in Fisher information that arises from the recalculation of the relative forces in the new frame of reference, $\bq'$. 

The balancing loss occurs because the average relative force, or fitness, is always zero in the current frame of reference, $\bq\cdot\ba=\sum q_i (\oq_i/q_i)=0$. Any direct gain in relative fitness by direct forces, $\obq\cdot\bF=\F$, must be balanced by an equivalent loss in relative fitness, $\obq\cdot\bI=-\F$, from the changing frame of reference in which relative fitness is calculated.

The movement of probability distributions in the canonical Price equation is always conservative, $\dd\abar=0$, so that d'Alembert's principle holds. When we transform to the general Price equation by $\ba\mapsto\bz$, then it may be that $\dd\zbar\ne0$ and the system is not conservative  \autocite{frank18the-price,frank17universal,frank15dalemberts}. In that case, we may consider constraints on $\dd\zbar$ and how those constraints influence the possible paths of change for $\dbq$. 

\subsection{Interpretation of force}

I have equated force with change. For example, $\bF=\ba=\dlog\bq$. The duality of force and change arises from the following relation. Given the initial condition and the force that acts on a population, we can deduce frequency change. Given the initial condition and the frequency change, we can induce the force \autocite{frank09natural}. 

The deduce-induce relation arises from the notion that force causes change. However, in the abstract mathematics, we only have the relation between force and change. The mathematics does not express primacy of one over the other.

The value of the abstract Price equation arises from its purely mathematical nature. By equating force with relative frequency change, we intentionally blur the distinction between external causes and internal effects. By describing change as the difference between two abstract sets rather than change through time or space, we intentionally blur the scale of change. By separating frequencies, $\bq$, from property values, $\bz$, we intentionally distinguish universal aspects of change between sets from the particular interpretations of property values in each application. 

The blurring of cause, effect and scale, and the separation of frequency from value, lead to abstract mathematical expressions that reveal the common underlying structure among seemingly different subjects. 

\section{Value of the partition}

The conservation of total probability and the constancy of relative success are by themselves trivial. So one might say that the Price equation is simply some notation to describe trivial facts. 

However, many fundamental equations from different disciplines follow immediately and easily from the Price equation partition. It seems that each discipline has, in its own way, come to the same essential invariant geometry of an underlying conservative system. 

Interpretation in different disciplines reduces to two aspects. First, one must separate the forces of direct interest from those other forces that alter the frame of reference. Second, one must distinguish the underlying conservative foundation from the coordinates of property values for particular problems. 

The Price equation does exactly and only those two aspects on which interpretation depends. By focus on those essential aspects, the Price equation brings out the unity of analysis between seemingly different subjects.

My prior publications have shown how key results from different disciplines arise simply and naturally from the Price equation \autocite{frank18the-price,frank17universal,frank15dalemberts}. I have already described a generalization of d'Alembert's principle of physical mechanics. The remainder of this article briefly sketches two additional examples. 

\section{Information}

The Price equation separates frequencies from property values. That separation shadows Shannon's separation of the information in a message, expressed by frequencies of symbols in sets, from the meaning of a message, expressed by the properties associated with the message symbols.  Price \autocite{price95the-nature} was clearly influenced by the information theory separation between frequency and property in his discussion of a generalized notion of natural selection that might unify disparate subjects. 

With regard to frequencies, the Price equation simply describes the universal expression of divergence between sets. By contrast, information theory interprets frequencies and changes in frequencies in terms of the information content of messages. 

What is the relation between the general, abstract Price equation description of frequencies in relation to the conservation of total probability and the information theory interpretation of frequencies as having some deeper meaning in terms a concept of ``information''?

I begin with the Price equation, which has no notion of ``information''. I show that key quantities and classic expressions of information theory follow immediately from the Price equation. I then consider the following question. 

Given that key expressions and results of information theory follow from the abstract Price equation, should we think of those results as deriving from information theory or from the expression of more basic principles of invariant geometry that arise solely from the conservation of total probability? I argue in favor of the second interpretation.

\subsection{Information expressions from the Price equation}

I showed in \Eq{fisherDist} that the first term of the Price equation is
\begin{equation*}
 \obq\cdot\bF=\obq\cdot\ba=\norm{\frac{\obq}{\sqrt{\bq}}}^2=\F,
\end{equation*}
in which $\F$ is an abstract, nondimensional expression of the Fisher information distance metric. The second term of the Price equation is
\begin{equation*}
  \dbq\cdot\bI=\bq\cdot\ba=-\F.
\end{equation*}
Thus, d'Alembert's principle for the Price equation
\begin{equation*}
  \lr{\bF+\bI}\cdot\dbq=\F-\F=0
\end{equation*}
expresses the conservation of total information. Fisher information \autocite{frieden04science} has occasionally been raised as a candidate for a fundamental principle underlying physics. 

By the Price equation, we see Fisher information and the conservation of information arising as simple consequences of the conservation of total probability. There is no essential need for an underlying notion of ``information''. 

The information theory interpretation can be very useful. The point here is to understand the underlying assumptions and mathematics that lead to such expressions.

The Kullback-Leibler divergence \autocite{kullback59information,kullback51on-information} is another key expression of information theory
\begin{align*}
  \KL{\bq'}{\bq}&=\sum_i q_i'\log\frac{q_i'}{q_i}=\bq'\cdot\dlog\bq\\[5pt]
  \KL{\bq}{\bq'}&=\sum_i q_i\log\frac{q_i}{q_i'}=-\bq\cdot\dlog\bq.
\end{align*}
This divergence measures the separation between two probability distributions, $\bq$ and $\bq'$. The Kullback-Leibler divergence provides an equivalent expression of Shannon information when the divergence is taken from an initial uniform distribution. Thus, the Kullback-Leibler divergence is often described as relative information---the change in information relative to some given initial distribution.

In information theory, it is often useful to consider the sum of the forward and backward divergences, which creates a symmetric measure. That sum is known as the Jeffreys divergence
\begin{align*}
  \J=\KL{\bq'}{\bq}+\KL{\bq}{\bq'}	&=\dbq\cdot\dlog\bq\\
  									&=\dbq\cdot\ba=\F.
\end{align*}
These results follow when changes are small. For analysis of discrete changes, see Frank \autocite{frank18the-price,frank17universal}. 

Note that, in biology
\begin{equation}\label{eq:malthus}
    \log\frac{q_i'}{q_i}=\log w_i=m_i,
\end{equation}
in which log fitness, $\log w_i=m_i$ is often called the Malthusian parameter. The information measures $\D$, $\J$ and $\F$ can all be expressed in terms of the Malthusian parameter.

\subsection{The interpretation of information}

All of the ``information'' results in the prior section arose directly from the canonical Price equation's description of conserved total probability. No notion or interpretation of ``information'' is necessary. 

In many disciplines, information expressions arise in the analysis of the specific disciplinary problems. This sometimes leads to the idea that information must be a primary general concept that gives form to and explains the particular results. 

Here, the Price equation explains why those information expressions arise so often. Those expressions are simply the fundamental descriptions of force and change within the context of a conserved total quantity. In this case, the conserved total quantity is total probability. 

Information does have many useful interpretations \autocite{cover91elements}. The next section provides an example.

\section{Inference: data as a force}

Following Bayesian tradition, denote the force of the data as $\tilde{L}(D|\Gth)$, the likelihood of observing the data, $D$, given a value for the parameter, $\Gth$. To interpret a force as equivalent to relative fitness, the average value of the force must be one to satisfy the conservation of total probability. Thus, define
\begin{equation*}
  w_\Gth=L_\Gth=\frac{\tilde{L}(D|\Gth)}{\sum_\Gth \qt\tilde{L}(D|\Gth)}.
\end{equation*}

We can now write the classic expression for Bayesian updating of a prior, $\qt$, driven by the force of new data, $L_\Gth=L(D|\Gth)$, to yield the posterior, $\qt'$, as
\begin{equation}\label{eq:bayesUpdate}
  \qt'=\qt L_\Gth.
\end{equation}

By recognizing $\bL$ as a force vector acting on frequency change, we can use all of the general results derived from the Price equation. For example, the Malthusian parameter of \Eq{malthus} relates to the log-likelihood as
\begin{equation}\label{eq:logL}
  \bmm=\log\frac{\bq'}{\bq}=\GD\log\bq=\log\bL.
\end{equation}

This equivalence for log-likelihood relates frequency change to the Kullback--Leibler expressions for the change in information
\begin{equation}\label{eq:bayesJ}
  \GD\bq\cdot\log\bL=\KL{\bq'}{\bq}+\KL{\bq}{\bq'},
\end{equation}
which we may think of as the gain of information from the force of the data. Perhaps the most general expression of change describes the relative separation within the unitary square root coordinates as the Euclidean length
\begin{equation*}
  \GD\bq\cdot\bL=\norm{\frac{\GD\bq}{\sqrt{\bq}}}^2,
\end{equation*}
which is an abstract, nondimensional expression for the work done by the displacement of the frequencies, $\GD\bq$, in relation to the force of the data, $\bL$. 

I defined $\bL$ as a normalized form of the likelihood, $\skew{-4}\tilde{\bL}$, such that the average value is one, $\skew{-4}\bar{\bL}=\bq\cdot\bL=1$. Thus, we have a canonical form of the Price equation for normalized likelihood
\begin{equation}\label{eq:canonicalL}
  \GD\skew{-4}\bar{\bL}=\GD\bq\cdot\bL+\bq'\cdot\GD\bL=0.
\end{equation}

The second terms show how the inertial forces alter the frame of reference that determines the normalization of the likelihoods, $\skew{-4}\tilde{\bL}\mapsto\bL$. Typically, as information is gained from data, the normalizing force of the frame of reference reduces the force of the same data in subsequent updates. 

All of this simply shows that Bayesian updating describes the change in probability distributions between two sets. That change between sets follows the universal principles given by the abstract Price equation. 

Prior work noted the analogy between natural selection and Bayesian updating \autocite{shalizi09dynamics,harper10the-replicator,campbell16universal}. Here, I emphasized a more general perspective that includes natural selection and Bayesian updating as examples of the common invariances and geometry that unify many topics. 

\section{Discussion}

How useful is it to understand the common mathematical basis of different disciplines? The generality by itself does not alter the well known results within each discipline. However, three points suggest potential value.

\medskip\noindent\textbf{Universality.} First, we can evaluate other claims of universality more clearly. For example, information expressions often occur in the fundamental equations of different disciplines. That commonality tempts one to think of information as a primary quantity. 

The Price equation shows that information expressions arise from the basic geometry of divergence between populations. Invariance and geometry set the universal foundations.

\medskip\noindent\textbf{Force and cause.} Second, I support Lanczos' \autocite{lanczos86the-variational} advocacy of d'Alembert's principle because ``it focuses attention on the forces, not on the moving body.'' The principle highlights causal interpretation of the forces that shape complex dynamics. 

D'Alembert's principle transforms a changing system into an equilibrium system. The direct forces by themselves cause nonequilibrium dynamics. The addition of the inertial forces brings the total system into equilibrium, $\lr{\bF+\bI}\cdot\dbq=0$.

Lanczos emphasizes that
\begin{quote}
By this device \textit{dynamics is reduced to statics}.

This does not mean that we can actually \textit{solve} a dynamical problem by statical methods. The resulting equations are \textit{differential equations} which have to be solved. We have merely \textit{deduced} these differential equations by statical considerations. The addition of the force of inertia $\bI$ to the acting force $\bF$ changes the problem of motion to a problem of equilibrium.
\end{quote}
This quote describes a great benefit of conservation laws in analysis.

In modern physics, d'Alembert's principle is often relegated to a historical footnote. The principle applies only to conservative mechanical systems. Most mechanical systems are not conservative, because they have forces, such as friction, that prevent changes from being reversed without loss. Because real systems are rarely conservative and reversible, d'Alembert's principle is limited even within its primary domain of mechanics. 

These criticisms of d'Alembert are true. But they miss the abstract mathematical power and insight of d'Alembert's expression, emphasized by Lanczos. Much of the great advance of modern physics came from the abstract structure of conservation laws and their invariances, which often have the balancing form of d'Alembert.

The canonical Price equation is a pure abstract expression of d'Alembert's balance between direct and inertial forces. In the canonical Price equation, with focus on frequencies and the change in probability distributions, the abstract system is frictionless, conservative, and reversible. Thus, the Price equation expresses the underlying mathematical structure that unifies so many seemingly different fundamental results of distinct disciplines, which share the same conservation of total probability or a similar conservation. 

For example, the equations of change by natural selection can initially be described in terms of changes in relative fitness. Because relative fitness concerns only frequencies, it matches the canonical Price equation, with a total change in relative fitness of zero. Within that equilibrium system, we can partition the change into the direct forces of natural selection and the changing frame of reference caused by altered frequencies. 

Typically, one is interested in the change in some trait value of organisms, $\bz$, rather than in the the change of relative fitness, $\ba$. Because $\bz$ is not conserved in the same way as $\ba$, additional forces may come into play. We can address those additional forces by a variety of powerful supplemental analytical methods, which include Jaynesian maximum entropy as a special case \autocite{frank17universal,frank18the-price}. However, there can never be a complete universal expression that captures all aspects of nonconservative systems.

\medskip\noindent\textbf{Particular vs general.} Third, the Price equation reveals that many particular explanations in science derive from general underlying principles. 

Physical work and information are well defined quantities with very useful applications. Yet the basic equations of change often arise from a general conservation law rather than from the particular qualities of work and information. 

To understand the fundamental results for physical work and information, one has to see the underlying invariant structure. Otherwise, one is naturally inclined to favor explanations in terms of particular physical or informational properties.

We certainly do not need the Price equation to understand work or information or the generic forms of probability distributions. But we are much more likely to see the fallacy of the particular if we begin from the Price equation perspective. The Price equation focuses attention on the invariant generic structure, not on the particular details.

In summary, the Price equation expresses the conservation of total probability. That conservation law constrains the dynamics of populations. The invariant geometry of change revealed by the Price equation explains why fundamental results in different disciplines often share the same underlying form.

\section*{Acknowledgments}

\noindent The Donald Bren Foundation supports my research. I completed this work while on sabbatical in the Theoretical Biology group of the Institute for Integrative Biology at ETH Zürich.


\mybiblio	


\end{document}